# BIDIRECTIONAL GROWTH BASED MINING AND CYCLIC BEHAVIOUR ANALYSIS OF WEB SEQUENTIAL PATTERNS

Srikantaiah K C[1], Krishna Kumar N[1], Venugopal K R[1], L M Patnaik[2]

[1]Department of Computer Science and Engineering,
University Visvesvaraya College of Engineering, Bangalore University, Bangalore,
`srikantaiahkc@gmail.com`
[2]Honorary Professor, Indian Institute of Science
Bangalore, INDIA

## ABSTRACT

*Web sequential patterns are important for analyzing and understanding users' behaviour to improve the quality of service offered by the World Wide Web. Web Prefetching is one such technique that utilizes prefetching rules derived through Cyclic Model Analysis of the mined Web sequential patterns. The more accurate the prediction and more satisfying the results of prefetching if we use a highly efficient and scalable mining technique such as the Bidirectional Growth based Directed Acyclic Graph. In this paper, we propose a novel algorithm called Bidirectional Growth based mining Cyclic behavior Analysis of web sequential Patterns (BGCAP) that effectively combines these strategies to generate prefetching rules in the form of 2-sequence patterns with Periodicity and threshold of Cyclic Behaviour that can be utilized to effectively prefetch Web pages, thus reducing the users' perceived latency. As BGCAP is based on Bidirectional pattern growth, it performs only (log n+1) levels of recursion for mining n Web sequential patterns. Our experimental results show that prefetching rules generated using BGCAP is 5-10% faster for different data sizes and 10-15% faster for a fixed data size than TD-Mine. In addition, BGCAP generates about 5-15% more prefetching rules than TD-Mine.*

## 1. INTRODUCTION

Data Mining is the process of extracting useful information from a large repository of data. Web mining is one of the types of data mining and can be defined as the process of discovery and analysis of useful information from the data corresponding to World Wide Web. There are three major types of Web Mining: (i) Web Structure Mining, (ii) Web Content Mining and (iii) Web Usage Mining.

Web Structure Mining is the process of using graph theory to analyze a Web graph where the nodes represent Web pages and the edges represent the hyperlinks among them. According to the type of web structural data, web structure mining can be divided into two categories: (i) extracting patterns from hyperlinks in the web and (ii) mining the document structure *i.e.,* analysis of the DOM tree structure of the pages to describe HTML or XML tag usage.

Web Content Mining is the process of mining actual content (text, image or multimedia) from the Web pages of the World Wide Web for information. Web content mining can be divided into two broad categories:

- **Agent-Based Approaches**

  Agent-based Web mining systems can further be divided into the following three categories:
  (i) *Intelligent Search Agents:* These Web agents are tools that interact with and learn the structure of unfamiliar Web pages and retrieve information from a variety of such sites using only general information about the domain.
  (ii) *Information Filtering/Categorization Tools:* A number of Web agents use various information retrieval techniques and characteristics of open hypertext Web documents to automatically retrieve, filter, and categorize them. Criteria for categorization can be semantic information embedded in link structures and document content to create cluster hierarchies of hypertext documents, and structure an information space.
  (iii) *Personalized Web Agents:* This category of Web agents learn user preferences and discover Web information sources based on these preferences and those of other individuals with similar interests.

- **Database Approaches**

  Database approaches in Web content mining focus on techniques for organizing the semi-structured data on the Web into more structured collections of resources, and using standard database querying mechanisms and data mining techniques to analyze it. There are two techniques in this approach:
  (i) *Multilevel Database Systems:* The main idea behind this technique is that the lowest level of the database contains semi-structured information stored in various Web repositories, such as hypertext documents. At the higher level, meta data or generalizations are extracted from lower levels and organized in structured collections, *i.e.,* relational or page-oriented databases.
  (ii) *Web Query Systems:* Web-based query systems utilize standard database query languages such as SQL for structural information about Web documents.

  Web Usage Mining is the automatic discovery of user access patterns from Web logs stored in different servers. Servers collect large volumes of data in their daily operations, generated automatically and collected in server access logs. Other sources of user information include referrer logs which contain information about the referring pages for each page reference, and user registration or survey data gathered via scripts. Analyzing such data can help in understanding the users' behaviour so that the server can improve its services like recommendation and Web personalization. Most Web analysis tools provide mechanisms for reporting user activity in the servers and various forms of data filtering. Using such tools it is possible to determine the number of accesses to the server and to individual files, the times of visits, and the domain names and URLs of users. These tools can be placed into two main categories such as:

- **Pattern Discovery Tools**
  These tools discover association rules and sequential patterns from server access logs. Sequential access patterns are essential for understanding and predicting the users' behaviour.

- **Pattern Analysis Tools**
  Once Web access patterns have been discovered, they need to be understood, visualized, and interpreted so that the knowledge gained by the analysis can be utilized to improve the services offered by the Web servers.

*A.  Motivation*

The Internet is an extremely large collection of network of networks, which in turn consist of Web servers which contain huge quantities of data, clients or end-user system which request information or services from the servers and finally client-side and server-side proxies which are additional systems that help and provide better communication amongst clients and servers. Such factors give rise to the necessity of creating intelligent systems that can effectively mine and analyze patterns and Web Usage Mining is one such technique. Mining from the Web includes integrating various data sources such as server access logs, referrer logs, user registration or profile information; and the importance of identifying user sessions or transactions from usage data, site topologies, and models of user behaviour. Also, analyzing such mined information results in predicting the users' behaviour and this knowledge in the form of prefetching rules is extremely helpful in reducing user's perceived latency and improving the quality of Web services.

*B. Contribution*

In this paper, we propose the Bidirectional Growth based mining and Cyclic behaviour Analysis of web sequential Patterns (BGCAP) Algorithm that generates Web prefetching rules. In our approach, we first perform preprocessing upon the raw Web logs to generate the session database of each Web user, which lists the Web users (IPs) along with their corresponding sessions. Then, we utilize a bidirectional pattern growth algorithm called UpDown Directed Acyclic Graph (UDDAG) [1] to generate Sequential Web access patterns from this session database and analyse them using Cyclic Model Analysis [2] to find out the Periodicity and Cyclic Behaviour of the mined 2-sequence patterns. The cyclic behaviour analysis can be used to generate Web prefetching rules. Constrains on Sequential Pattern Mining like date and time are specified, so that it returns interesting, more desirable, useful navigational patterns instead of huge and unwanted patterns.

*C. Organization*

The remainder of the paper is organized as follows. In the next section the related works and existing techniques are discussed briefly. In Section 3, we explain the background of this paper. Section 4 contains the problem definitions, assumptions and basic definitions. In Section 5, we discuss the system model and the BGCAP Algorithm and the architecture. Section 6 provides the results and the performance of BGCAP. Finally, Section 7 contains conclusions.

## 2. Related Work

Several techniques have been proposed for sequential pattern mining. They are mainly of two types: (i) *Apriori* based (ii) Frequent Pattern growth and (FP-growth) based. *Apriori* based mining techniques such as *Apriori*-all, *GSP* [3], SPADE [4], LAPIN-SPAM [5], LAPIN [6], scan the database multiple times. A *n* size pattern requires *n* scans of the database and hence these mining techniques are generally inefficient. FP growth based mining techniques such as FreeSpan, BIDE [7], COBRA [8], PrefixSpan [9], UDDAG [1], etc., utilize a tree based representation that reflects the original database and two scans are required to construct the tree. From this tree, the sequential patterns are derived without reference to the original database. Changes in the original database can easily be reflected in the tree by incremental analysis. These sequential pattern mining algorithms are used for mining Web access patterns from Web logs. Their variants are summarized as follows:

Cheng, et al., [10] proposed an approach that combines *Apriori*-all and clustering for sequential pattern mining to identify the user pattern and cluster users' path patterns and make the similar users cluster as one group in order to reduce the pattern that is effective for users in personalized service. But, it should be taken into account that not all patterns in a cluster may prove to be useful as they can produce erroneous conclusions after pattern analysis.

Gaol [11] explored habits of users using *Apriori*-all algorithm which first stores the original web access sequence database for storing non-sequential data. This is based on the fact that the greater the number of combinations produced, the less likely the number of users who perform a combination of these and vice versa. While such an approach is simple and straight-forward, such brute force tactics are obsolete as *Apriori*-all algorithms are found to be least efficient with respect to sequential pattern mining.

Pei, et al., [12] proposed Web Access Pattern tree (WAP-tree) for efficient mining of access patterns from Web logs. The Web access pattern tree stores highly compressed, critical information for sequential pattern mining. The WAP-tree registers all access sequence counts. There is no need for mining the original database any more as the mining process for all Web access patterns needs to work on the WAP-tree only. Therefore, WAP-mine needs to scan the access sequence database only twice. The height of the WAP-tree is one plus the maximum length of the frequent subsequences in the database. The width of the WAP-tree, *i.e.,* the number of leaves of the tree, is the number of access sequences in the database. The size of the WAP-tree is much smaller than the size of access sequence database. It is shown that WAP-mine outperforms and has better scalability than GSP.

Xiaoqiu, et al., [13] proposed the Improved WAP-tree in the form of highly compressed access sequences and introducing a sub-tree structure to avoid generation of conditional WAP tree repeatedly and to generate maximal sequences. Improved WAP-tree excels traditional WAP-tree in time and space, and shows better stability as the lengths of patterns vary. Also, mining frequent access sequences based on WAP-tree needs to scan transaction database only twice.
Yang, et al., [14] designed an efficient algorithm Top Down Mine (TD-mine) which makes use of the WAP tree data structure for web access pattern mining. WAP tree can be traversed both top-down and bottom-up for the extraction of frequent access patterns. In TD-mine, a header table is used to traverse the tree from the root to the leaf nodes and mine patterns where the nodes are frequently accessed.

Liu, et al., [15] proposed the Breadth-First Linked WAP-tree (BFWAP-tree) to mine frequent sequences which reflects parent-child relationship of nodes. The proposed algorithm builds the frequent header node links of the original WAP-tree in a Breadth-First fashion and uses the layer code of each node to identify the parent-child relationships between nodes of the tree. It then finds each frequent sequential pattern through progressive Breadth-First sequence search, starting with its first Breadth-First subsequence event. BFWAP avoids re-constructing WAP-tree recursively and shows a significant performance gain.

Vijayalakshmi, et al., [16] designed an extended version of PrefixSpan called EXT-Prefixspan algorithm to extract the Constraint-based multidimensional frequent sequential patterns in web usage mining by filtering the dataset in the presence of various pattern constraints. EXT-PrefixSpan then mines the complete set of patterns but greatly reduces the efforts of candidate subsequence generation. This substantially reduces the size of projected database and leads to efficient processing. EXT-PrefixSpan can be used to mine frequent sequential patterns of multi-dimensional nature from any web server log file in the light of obtaining the frequent web access patterns. However, EXT-PrefixSpan does not specify any particular constraint for consideration *i.e.,* it is highly generic.

Wu, et al., [17] proposed the CIC-PrefixSpan, a modified version of PrefixSpan that mines and generates Maximal Sequential patterns by combining PrefixSpan and pseudo-projection. First, preprocessing is done to categorize the user sessions into human user sessions, crawler sessions and resource-download user sessions for efficient Web sequential pattern mining by filtering out the non-human user sessions, leaving the human user sessions and finding the transactions using Maximum Forward Path (MFP). By utilizing CIC-PrefixSpan, the memory space is reduced and generating duplicate projections to find the most frequent path in the users' access path tree is also avoided. It is shown that CIC-PrefixSpan yields accurate patterns with high efficiency and low execution time compared to GSP and PrefixSpan. However, the frequent substructures within a pattern cannot be mined by CIC-PrefixSpan.

Verma et al., [18] designed a new pattern mining algorithm called Single Level Algorithm for extracting behaviour patterns. These patterns are used to generate recommendations at run time for web users. Single Level Algorithm is designed keeping in mind the dynamic adaptation of *focused* websites that have a large number of webpages. Single It combines preprocessing, mining and analysis to eventually predict the users' behaviour and hence is useful for specific websites and is highly scalable. It is shown to be more efficient than *Apriori* algorithm. However, performing preprocessing on a very large Web log database can be time-consuming and too cumbersome to be integrated with mining and analysis.

Nasraoui, et al., [19] presented a framework for discovering and tracking evolving user profiles in real-time environment using Web usage mining and Web ontology. Preprocessing is first performed on the Web log data to identify user sessions. Then, profiles are constructed for each user and enriched with other domain-specific information facets that give a panoramic view of the discovered mass usage modes. This framework summarizes a group of users with similar access activities and consists of their viewed pages, search engine queries, and inquiring and inquired companies. By mapping some new sessions to persistent profiles and updating these profiles most sessions are eliminated from further analysis and focusing the mining on truly new sessions. However, this framework is not scalable.

Pitman, et al., [20] modified the Bi-Directional Extension (BIDE) algorithm for mining closed sequential patterns in order to identify domain-specific rule sets for recommendation of pages and personalization for web users in E-commerce. Individual supports are specified for each customer so that products can be recommended for individuals. Also, BIDE creates multidimensional sequences and further increase prediction for customers who do not explicitly specify their needs by using search functionality. However, additional strategies must be explored for identifying the most relevant sequential patterns without an exhaustive exploration of the search space bounded only by minimum support.

Masseglia, et al., [21] proposed a Heuristic based Distributed Miner (HDM), a method that allows finding frequent behavioural patterns in real time irrespective of the number of web users. Navigational schemas, that are completely adaptable to the changing Web log data, are provided by HDM for efficient frequent sequence pattern mining. Based on a distributed heuristic, these schemas provide solutions for problems such as (i) discovering *interesting zones* (a great number of frequent patterns concentrated over a period of time) (ii) discovering *super-frequent* patterns and (iii) discovering very long sequential patterns and interactive data mining. However, the quality of the schemas can further be improved by adapting the candidate population.

Zhou, et al., [22] designed an intelligent web recommender system known as Sequential Web Access based Recommender System (SWARS) for sequential access pattern mining. Conditional Sequence mining (CS-mine) algorithm is used to identify frequent sequential web access patterns. The access patterns are then stored in a compact tree structure, called Pattern-tree, which is then used for matching and generating web links for recommendations. SWARS has shown to achieve good performance with high satisfaction and applicability.

Yen, et al., [23] address the issue of re-discovery of dynamic web logs due to the obsolete web logs as a result of deletion of users' log data and insertion of new logs. Incremental mining utilizes previous mining results and finds new patterns from the updated (inserted or deleted) part of the web logs. A new incremental mining strategy called Incremental Mining of Web Traversal Patterns (IncWTP) is proposed which makes use of an incremental updating algorithm to maintain the discovered path traversal patterns when entries are inserted or deleted in the database. This is achieved by making use of an extended lattice structure which is used to store the previous mining results. However, the changes made to the website structure will not be reflected in the lattice structure.

Zhang et al., [24] applied the Galois lattice to mine Web sequential access patterns by representing the paths traversed using graphs and compare the performance with that of *Apriori*. Since the Apriori-like algorithms frequently scan entire transaction database to generate candidate patterns, Galois lattice reduces time complexity of closed sequential pattern mining as it needs only one scan. Jain, et al., [25] proposed a technique that employs Doubly Linked Tree to mine Web Sequential patterns. The web access data available is constructed in the form of doubly linked tree. This tree keeps the critical mining related information in compressed format based on the frequent event count. It is shown that for low support threshold and for large data base Doubly Linked Tree mining performance is better than conventional schemes such as *Apriori*-all and GSP. However, Doubly Linked Tree does not work well in a distributed environment.

Jha, et al., [26] proposed a Frequent Sequential Traversal Pattern Mining based on dynamic Weights constraint of web access sequences (FSTPMW) to find the information gain of sequential patterns in session databases. The weight constraints are added into the sequential traversal pattern to control number of sequential patterns that can be generated in addition to minimum threshold. FSTPMW is efficient and scalable in mining sequential traversal patterns. But, it should be noted that FSTPMW does not consider levels of support along with the weights of sequential traversal patterns. Wang, et al., [27] proposed a Web personalization system that uses sequential access pattern mining based on CS-mine algorithm. The access patterns are stored in a compact tree structure called Pattern-tree which is then used for matching and generating web links for recommendations. Pattern tree has shown to achieve good performance with accurate predictability.

Saxena, et al., [28] integrated mining and analysis by proposing the One Pass Frequent Episode discovery (FED) algorithm. In this approach significant intervals for each website are computed first and these intervals are used for detecting frequent patterns (Episodes). Analysis is then performed to find frequent patterns which can be used to forecast the user's behavior. The FED algorithm is very efficient as it finds out patterns within one cycle of execution itself. Oikonomopoulou, et al., [29] proposed a prediction schema based on Markov Model that extracts sequential patterns from Web logs using web site topology. Full coverage is achieved by the schema while maintaining accuracy of the prediction. Since Markov Models are infamous for their precision, the proposed prediction schema fails to deploy a more complex categorization method for each sequential pattern.

Rajimol, et al., [30] proposed First Occurrence List Maximum (FOLMax-mine) for mining maximal web access patterns based on FOL-Mine. It is a top-down method that uses the concept of first occurrence to reduce search space and improve the performance. This is achieved by finding out the Maximal Frequent Path in the patterns generated from the Web logs.

## 3. BACKGROUND

WAP-Mine is one of the FP-growth based algorithms for mining frequent web access patterns from web access database. But in the process of mining frequent web access patterns, WAP-Mine produces many intermediate data which brings down the efficiency especially at lower support. TD-Mine, which is an extension of WAP-Mine, overcomes this problem and saves more space by reducing the amount of intermediate data generated. But, TD-Mine needs $n$ levels of recursions to mine a pattern of length ($n$+1).

In this paper, we propose BRCAP Algorithm that utilizes UDDAG based on Bidirectional pattern-growth approach which in turn focuses the search on a restricted portion of initial database to avoid expensive candidate generation and test step and is better on mining longer patterns. The strategies used in UDDAG are partitioning, projection and detection. UDDAG only needs (log $n$+1) levels of recursion thereby significantly reducing the execution time compared to TD-Mine. It takes minimal processing power for mining the complete set of sequential patterns in a large sequence database. These sequential patterns are analyzed using Cyclic Model Analysis that deals with the tendency of certain sequential patterns to repeat themselves periodically after definite time intervals. This concept is greatly helpful in predicting future browsing patterns and prefetching Web pages aimed at certain user groups.

# 4. PROBLEM STATEMENT

## A. Problem Definition

Given a Web log database $W$ and the set of pages $P = \{p_i : 1<=i<=n\}$ in a Web server, a *session* is a subset of $P$, denoted by $(p_1, p_2,... p_k)$, where $p_i \in P$, $i \in \{1, …, k\}$. Here, the parentheses are omitted for a session with one page only. A *Web access sequence q* is a list of sessions, denoted by $< q_1\ q_2\ …\ q_m >$, where $q_i$ is a session and $q_i \subseteq P$, $i \in \{1, …, m\}$. The number of sessions in $q$ is called the length of $q$. Given two access sequences $x = < x_1\ x_2\ …\ x_j >$ and $y = < y_1\ y_2\ …\ y_k >$, $x$ is said to be a *subsequence* of $y$ and $y$ a *supersequence* of $x$ if $k \geq j$ and there exists integers $1 \leq i_1 < i_2 < … < i_j \leq k$, such that $x_1 \subseteq y_{i1}$, $x_2 \subseteq y_{i2}, …, x_j \subseteq y_{ij}$. Here, $x$ is also contained in $y$ which is denoted by $x \propto y$. A *session database* is a set of tuples $<ip, q>$, where $ip$ is the user's IP address which is used as a sequence identifier and $q$ is the user's access sequence. A tuple $<ip, q>$ is said to contain an access sequence $\alpha$ if $\alpha \propto q$.

The *Support* of a subsequence $\alpha$, denoted by *Support* $(\alpha)$, is the number of sequences for which $\alpha$ is a subsequence. A subsequence $\alpha$ is said to be a *Web sequential access pattern* when its support is greater than user-specified minimum support (*MinSup*) *i.e., Support* $(\alpha) \geq MinSup$.

Given a session database and *MinSup*, the objective is to extract Web sequential access patterns using Bidirectional Pattern Growth algorithm, analyze them using Cyclic Model Analysis to determine the periodicity and cyclic behaviour of the 2-sequence patterns and employ them as prefetching rules.

## B. Assumptions

During the preprocessing operation of the Web logs, it is assumed to be sufficient to consider only those URLs whose domains contain text and images only. Web sequential patterns pertaining to Multimedia Web pages need not be mined and hence are filtered out during preprocessing of the raw Web logs.

## C. Basic Definitions

**Web Prefetching:** Web Prefetching is the process of fetching Web pages from the Web server before they are actually requested by the users.

**Periodicity:** The periodicity $P$ of the 2-sequence $<p_i\ p_j>$ is defined as the time period $t$ after which $p_j$ shall be accessed periodically after $p_i$ has been accessed. Example: The periodicity $P$ between a pair of Web pages $(p_i, p_j)$ is 10 ms indicates that the page $p_j$ is accessed periodically with period 10 ms after page $p_i$ has been accessed.

**Tendency:** The tendency between a pair of Web pages $(p_i, p_j)$ is defined as the line in the trend graph that determines whether the cyclic behaviour of the 2-sequence $<p_i\ p_j>$ is increasing or decreasing.

**Cyclic Behaviour:** The cyclic behaviour $C$ of the 2-sequence $<p_i\ p_j>$ is defined as the time that gives the stopping criteria for accessing page $p_j$ after $p_i$ has been accessed. It is calculated using Periodicity and Tendency. The periodicity is increased by adding the value to itself each time the page is accessed. The page will not be accessed after periodicity has reached the limit of cyclic behaviour *i.e.,* $P \leq C$. Example: The periodicity $P$ between a pair of Web pages $(p_i, p_j)$ is 10 ms, the trend line is decreasing and the cyclic behaviour $C = 50$ ms indicates that accessing of page $p_j$ repeats for every 10 ms after accessing the page $p_i$. This ends when the time $t$ *reaches* 50 ms.

# 5. SYSTEM MODEL

*A. System Architecture*

The system architecture consists of the following components, (i) Web logs, (ii) Preprocessing Engine, (iii) Encoder, (iv) Sequential Pattern Miner, (v) Pattern Analyzer and (vi) Prefetching Rules is as shown in the Figure 1.

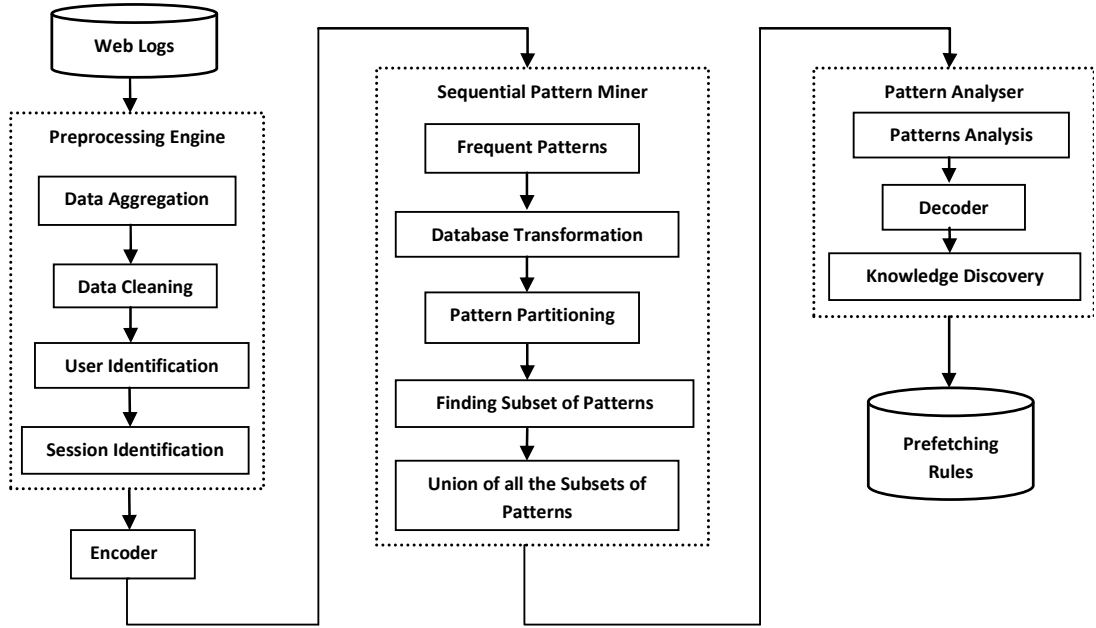

Figure 1: System Architecture

**Web Logs**

A Web log is a large database stored in Web servers that contains details of the transactions of Web users. There are many fields in the Web log database, which conform to either Common Log Format (CLF) or the Extended Common Log Format (ECLF) as shown in the Figures 2 and 3 respectively. The fields specified by this format are IP address of the destination page, Destination URL, Code which denotes the packet number, Protocol which specifies the type of Network protocol used (e.g. TCP), Method which specifies the type of method HTTP method used (e.g. GET, POST), Type which specifies the file type, Date which gives the date and time when the page was accessed, Referrer which gives the IP address of the client which requested the page, and finally the Size of the page in bytes. An extract from a Web log is as shown in Figure 4. Web logs are very useful for predicting the behaviour of the clients and hence different mining techniques can be used to find and extract interesting browsing patterns of the users.

```
<ip_addr><base_url><date><method><file><protocol><code><bytes>
```

Figure 2: Common Log Format (CLF)

```
<ip_addr><base_url><date><method><file><protocol><code><bytes><referrer>
```

Figure 3: Extended Common Log Format (ECLF)

```
web-proxy, debug, packet 1307775248.816    363 30.0.1.2 TCP_MISS/200 960 GET
http://www.facebook.com/ajax/typeahead/search.php? - DIRECT/66.220.146.32
application/x-javascript in 11-Jun 12:25:6.76 from 30.0.7.254
web-proxy, debug, packet 1307775249.609    586 30.0.0.223 TCP_MISS/200 397 POST
http://channel.tvunetworks.com/list/all - DIRECT/38.103.62.170 text/html in 11-Jun
12:25:7.69 from 30.0.7.254
```

Figure 4: Access log from the Web server

**Preprocessing Engine**

This component aggregates the Web logs data from different sources and produces session database as the result using the following steps:

- **Data Aggregation**

    The Web logs from different sources with different formats are extracted and integrated into a single database so that they pertain to a single format with no redundancy.

- **Data Cleaning**

    In this step the aggregated data is cleansed, *i.e.,* useless records such as URLs containing images, multimedia, scripts and entries corresponding to crawlers are removed and only human initiated entries (*i.e.,* URLs ending with HTM, HTML, XHTML, PHP and JSP) are retained. Only a few of these fields are important for the mining process and hence after extracting or collecting only the important fields such as the User IP Addresses (referrers), URLs, and Date and the type of file in the URL (whether text, image, or script), the rest can be ignored.

- **User Identification**

    The identity of a user is not a prerequisite for Web usage mining. However, it is necessary to differentiate users to identify sessions. The cleaned database is grouped by different IP addresses and sorted by Date and Time for each IP for identifying different users. As the database is used to store this information, simple queries can be used to achieve this operation. By doing so, we can identify individual users and also find out the total number of users.

- **Session Identification**

    Session Identification is the process of partitioning the user activity records of each user into sessions in order to reconstruct the actual sequence of actions done by each user. A session is a package of activities that consists of a user's navigation history. These sessions are then aggregated to create a session database. The user activity records are divided into sessions by assigning unique identifiers for each session. Each user is assigned a separate session and also a separate session is assigned for the same user if the user exceeds a certain threshold of time (e.g. 15 minutes).

**Encoder**

Encoding is the process where each URL in the preprocessed database is assigned a unique identifier which is a non-negative integer. This is done as it would be too cumbersome to mention the URLs by their domain names in the session database. For this purpose, a table consisting of the list of URLs can be maintained where they are mapped to their respective identifiers. User Identification, Session Identification and Encoding of URLs are all shown in the example in the Figure 5. After encoding, the session database, which consists of users and their sessions, is generated and an example is shown in Figure 6.

Figure 5: Preprocessed Data

Figure 6: Session Database

**Pattern Miner**

Pattern Mining is used to find hidden patterns from large database when a minimum threshold of occurrence (*MinSup*) has been specified. The Pattern Miner utilizes UDDAG that grows patterns bi-directionally along both ends of detected patterns and allows faster pattern growth with fewer levels of recursion. UDDAG eliminates unnecessary candidates and supports efficient pruning of invalid candidates. This represents a promising approach for applications involving searching in large spaces like Web Sequential Pattern Mining.

**Pattern Analyser**

Pattern Analysis is the process to study and conduct the analysis of the results obtained from the Web sequential access patterns derived by the miner. The analyzer utilizes CMA to find out the periodicity and cyclic behaviour of all the 2-sequence patterns mined. Decoding is the process where the identifiers are replaced with their corresponding URLs. In the Knowledge Discovery process, the analysed data is transformed into Web prefetching rules and sent to the Prefetching Rule Depository so that they can be used to prefetch and cache Web pages and reduce the round-trip delay experienced by the users.

**Prefetching Rule Depository**

This component is a large database consisting of prefetching rules pertaining to the requested Web pages. After the pattern analysis, each 2-sequence pattern having periodicity and cyclic behaviour are stored here in the form of prefetching rules and are triggered when the first page in the sequence is accessed by a user.

### B. Algorithm

Given a raw Web log database *W*, we first perform the preprocessing and generate the session database *SD* as shown in Table 1. The session database consists of a set of tuples, with each tuple consisting of the user (IP) and the access sequence of that user. From Table 1, the user with IP address 1.0.1.2 has accessed the Web pages A, B, C and D with the sequence <A (A,C) B D>, where {A, B, C, D, E, F} is the set of unique Web pages accessed by  different users. Here (A, C) in the above sequence denotes that in a single session, the two Web pages A and C were visited by the user in that order, otherwise the user is assumed to visit a single page per session if the parentheses () are not specified.

Table 1. Example Sessions Database.

| User (IP) | Sequences |
|-----------|-----------|
| 1.0.1.2 | \<A (A,C) B D\> |
| 1.0.1.3 | \<A D (E, F)\> |
| 1.0.1.4 | \<(B, D) C F\> |
| 1.0.1.5 | \<(C, E) (A, B, C, D) \> |
| 1.0.1.6 | \<A  B C D E\> |

Web sequential patterns are mined from the session database *SD* using updated version of UDDAG [1]. This mining technique is basically a divide-and-conquer approach which tries to construct the patterns simultaneously along both directions of a Directed Acyclic Graph (DAG). The approach consists of three main steps: (i) Database Transformation (ii) Pattern Partitioning and (iii) Finding subsets of Patterns.

Database Transformation is used to remove infrequent pages, *i.e.,* those pages that do not have *MinSup* = 2. From Table 1, the frequent items with *MinSup* $\geq$ 2 are (A), (B), (C), (D), (E), (F), (A, C), (B, D). Since we require only these patterns, the remaining can be eliminated by substituting these patterns with non-negative integers, like so: (A)-1, (A, C)-2, (B)-3, (B, D)-4, (C)-5, (D)-6, (E)-7, (F)-8. For the simplicity of representation, we assign each IP a unique identifier as well. The transformed database is as shown in the Table 2.

Table 2. After Database Transformation.

| User (IP) | Sequences |
|-----------|-----------|
| P | \<1 2 3 6\> |
| Q | \<1 6 (7, 8)\> |
| R | \<4 5 8\> |
| S | \<(5, 7) (1, 3, 5, 6) \> |
| T | \<1 3 5 6 7\> |

Next, the patterns have to be partitioned into projected databases for each pattern, denoted by $^nD$, where *n* represents the frequent item. By partitioning, we select only that tuples in which *n* is present, called projected database for item *n* and ignore the rest. So, Table 2 is partitioned into *m* projected databases as shown in Table 3.

Table 3. Projected Databases.

| $^1D$ | | $^2D$ |
|-------|-------|-------|
| **p:** \<1 2 3 6\>   **q:** \<1 6 (7, 8)\> **s:** \<(5, 7) (1, 3, 5, 6) \>  **t:** \<1 3 5 6 7\> | | **p:** \<1 2 3 6\> |
| $^3D$ | | $^4D$ |
| **p:** \<1 2 3 6\> **s:** \<(5, 7) (1, 3, 5, 6)\> **t:** \<1 3 5 6 7\> | | **r:** \<4 5 8\> |
| $^5D$ | | $^6D$ |
| **r:** \<4 5 8\> **s:** \<(5, 7) (1, 3, 5, 6) \>  **t:** \<1 3 5 6 7\> | | **p:** \<1 2 3 6\>   **q:** \<1 6 (7, 8)\> **s:** \<(5, 7) (1, 3, 5, 6)\>  **t:** \<1 3 5 6 7\> |
| $^7D$ | | $^8D$ |
| **q:** \<1 6 (7, 8)\> **s:** \<(5, 7) (1, 3, 5, 6)\> **t:** \<1 3 5 6 7\> | | **q:** \<1 6 (7, 8)\> **r:** \<4 5 8\> |

The third and final step, finding subsets of patterns is not as straightforward as the previous steps. If $WP$ is the set of all sequential patterns mined from $W$ with $MinSup = 2$, then let $WP_1$, $WP_2$, …, $WP_8$ be the subsets of patterns, *i.e.,* the pattern mined from $^1$D, $^2$D, …, $^8$D respectively. If the condition $|^nD| \geq 2$ is not met then such projected databases can be ignored. Here, the projected databases $^2$D and $^4$D can be ignored as they contain only one tuple each and hence do not contribute frequent patterns. So we now have to find $WP_1$, $WP_3$, $WP_5$, $WP_6$, $WP_7$ and $WP_8$.

Let us consider $^6$D. As seen in Table 3, the projected database for 6 contains four tuples with IPs $p$, $q$, $s$ and $t$. We have to find out the sequential patterns in $^6$D by recursively partitioning the projected database into prefix and suffix databases until no frequent patterns are found. In the first round of partitioning, we split the projected database into Prefix (Pre ($^6$D)) and Suffix (Suf ($^6$D)) subsets, each containing the tuples with the prefix and suffix sequences respectively, pertaining to 6 as shown in the Table 4:

Table 4. Prefixes and Suffixes for $^6$D.

| Sequence Sets | Frequent Patterns |
|---|---|
| **Pre ($^6$D):** {<1 2 3>, <1>, <(5, 7)>, <1 3 5>} | <1>, <3>, <5> |
| **Suf ($^6$D):** {<(7, 8)>, <7>} | <7> |

Again, the frequent items, *i.e.,* the patterns in Pre ($^6$D) are <1>, <3> and <5>, and the pattern in Suf ($^6$D) is <7>. So, Pre ($^6$D) is further split into $PP_1$, $PP_3$, and $PP_5$ and Suf ($^6$D) is split as $PS_7$ and this process continues recursively until no frequent patterns are available. The prefix and suffix databases pertaining to $PP_1$, $PP_3$, $PP_5$ and $PS_7$ are as shown in Table 5:

Table 5. Prefixes and Suffixes of Prefix and Suffix Databases of $^6$D.

| Sequence Sets | Frequent Patterns |
|---|---|
| **Pre ($PP_1$):** {<>} | – |
| **Suf ($PP_1$):** {<(2, 3)>, <(3, 5)>} | <3> |
| **Pre ($PP_3$):** {<1 2 >, <1>} | <1> |
| **Suf ($PP_3$):** {<5>} | – |
| **Pre ($PP_5$):** {<1 3>} | – |
| **Suf ($PP_5$):** {<>} | – |
| **Pre ($PS_7$):** {<>} | – |
| **Suf ($PS_7$):** {<>} | – |

From Table 5, it can be seen that only two patterns <3> and <1> are frequent and no other frequent patterns exist in projected database for <3> and <1>. Hence, the partitioning stops and the DAG is constructed as shown in the Figure 7 to find all the patterns of $^6$D.

In Figure 7, <6> is the root node of the DAG and it contains occurrence set {p q s t} to show that the pattern <6> occurs in the tuples with IPs $p$, $q$, $s$ and $t$. In a DAG, the Up-children and Down-children denote the prefixes and suffixes of the root node respectively. Here, the root node <6> has three Up-children and one Down-child. Pre ($^6$D) yielded <1>, <3> and <5> and hence these are prefixes of <6> and <1 6>, <3 6> and <5 6> are frequent patterns as they occur in tuples with IPs {p q t}, {p t} and {s t} respectively. Suf ($^6$D) yielded <7> and is a suffix of <6> and hence <6 7> is a frequent pattern as it occurs in tuples with IPs {q t}. From Suf ($PP_1$), suffix of <1> is <3> and from Pre ($PP_3$), the prefix of <3> is <1>, therefore, <1 3 6>, the upmost child node, is a frequent pattern as it occurs in tuples with IPs {p t}. The patterns in an Up-child and Down-child of a DAG can be combined to form a new node only if the number of tuples of the occurrence sets of the Up-child and Down-child is greater than or equal to $MinSup$. So, the node <1 6 7> has one Up-parent <1 6> and one Down-parent <6 7> and as it is a frequent pattern in {q t}, <1 6 7> is a valid sequential pattern.

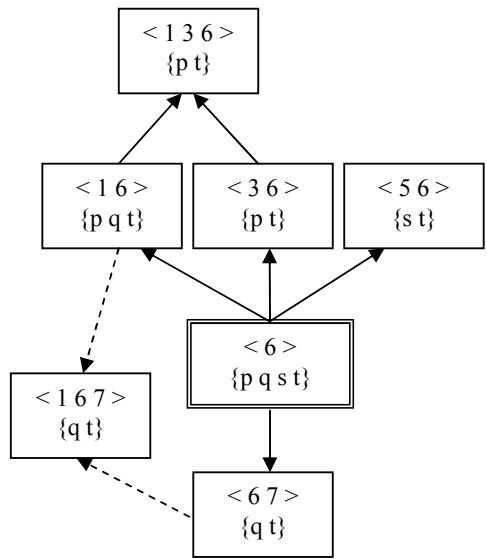

**Figure 7: The Example DAG**

Therefore, the complete set of patterns in $^6$D is, $WP_6 = \{<6>, <1\ 6>, <3\ 6>, <5\ 6>, <6\ 7>, <1\ 3\ 6>, <1\ 6\ 7>\}$. Similarly, $WP_1 = \{<1>, <1\ 3>, <1\ 6>, <1\ 3\ 6>, <1\ 6\ 7>\}$, $WP_3 = \{<3>, <1\ 3>, <3\ 6>, <1\ 3\ 6>\}$, $WP_5 = \{<5>, <5\ 6>\}$, $WP_7 = \{<7>, <1\ 7>, <6\ 7>, <1\ 6\ 7>\}$ and $WP_8 = \{<8>\}$. The complete set of patterns in the session database is $WP = \{<1>, <3>, <5>, <6>, <7>, <8>, <1\ 3>, <1\ 6>, <1\ 7>, <3\ 6>, <5\ 6>, <6\ 7>, <1\ 3\ 6>, <1\ 6\ 7>\}$. Maximal forward references save memory and can be used to generate all the above sequential patterns for WP and so WP (max) is $\{<8>, <5\ 6>, <1\ 3\ 6>, <1\ 6\ 7>\}$. As this paper focuses on deriving prefetching rules, we use WP (max) to generate all 2-sequence Web access patterns. Set of all 2-sequence patterns are $\{<1\ 3>, <1\ 6>, <3\ 6>, <1\ 7>, <6\ 7>, <5\ 6>\}$. These 2-sequence patterns are analyzed using Cyclic Model Analysis (CMA) [2] to find out the periodicity and cyclic behaviour of these sequences in the sequence database $D$. After analysis, the set of prefetching rules $PR$ are derived from the periodicity and cyclic behaviour and stored in the server.

**Table 6. BGCAP Algorithm.**

```
//Purpose:  To find Periodicity and    Tendency of sequential
               patterns.
//Input:    W (Web log Database)
//Output:   PR (Set of Prefetching Rules)
Begin
    F = Cleaning(W)
    SD = SessionIdentifier(F,THRESHOLD)
    WP = Bidirection Pattern GrowthP(SD,minsup)
    PR = Pattern Analysis(WP)
End
Cleaning(W)
Begin
    for each l ∈ W do
        if(URL in l contains (js, css))

        else
            insert l into F
        end if
    end for
End
```

```
SessionIdentifier(F,THRESHOLD)
//Purpose: To identify Sessions
// Input: F- Cleaned database and sorted web log entries according
           to IP address and time, Threshold -time limit for each
           session (L_i.ip - IP address at record L_i and
           L_i.t- Date/Time entry at record L_i )
//Output: F- with Sessions
Begin
    session_id = 0
    for each l ∈ f d
        if(l_i.ip != l_{i+1}.ip) then
            session_id++
            l_i.sid = session_id
        else if(l_i.ip == l_{i+1}.ip and (l_{i+1}.t - l_i.t) > threshold) then
            session_id++
            l_i.sid = session_id
        else
            l_i.sid = session_id
        end if
    end for
End
```

For example, if the 2-sequence pattern <3 6> occurs frequently in *D* with a Periodicity of 10 seconds, that means the user accesses <6> 10 seconds after he has accessed <3>. If this behaviour repeats itself and then stops after 80 seconds, then it is said to be the threshold of Cyclic Behaviour of <3 6> after which this pattern will not repeat again. Using this knowledge, the Web page <6> can be prefetched for the user before the 10th second and stored in the cache for future references and hence reduce his perceived latency. The Web page can be deleted from the cache when the Cyclic Behaviour of 80 seconds has been reached. The algorithm summarizes the above processes and is shown in the Table 6.

## 6. EXPERIMENTAL RESULTS

The algorithm BGCAP has been implemented using Java language using Netbeans 6.9.1 platform on MSNBC dataset in a Pentium Dual Core processor environment, with a 2GB Memory and 100 GB HDD. The MSNBC Web log data comes from Internet Information Server (IIS) logs for msnbc.com and news-related portions of *msn.com* for 989818 users. Each sequence in the dataset corresponds to page views of a user during that twenty-four hour period. Each event in the sequence corresponds to a user's request for a page. The page requests served *via* caching mechanism are not recorded in the server logs and hence, not present in the data.

Experiments are pursued to compare the efficiency of BGCAP and TD-Mine. BGCAP demonstrated satisfactory scale-up properties with respect to various parameters such as the total number of Web access sequences, the total number of pages, the average lengths of sequences. The following comparisons show that BGCAP outperforms TD-Mine in quite a significant margin and has better scalability than TD-Mine.

The data size is the number of transactions of the input session database and it is a significant factor that affects the performance of BGCAP. This is demonstrated in Figure 8 where we show how different data sizes have different execution times. The higher the number of transactions, the more time it would take to process the data and generate patterns. However, as seen in Figure 8, TD-Mine's execution time is moderately higher than that of BGCAP. This is because of the Bidirectional pattern-growth approach adopted by BGCAP that reduces the run time by 5-10% than that of TD-Mine. However, as the data size increases to the order of about a million transactions, the run time of both the approaches will tend to be the same since BGCAP must recursively generate prefixes and suffixes for all the frequent items mined.

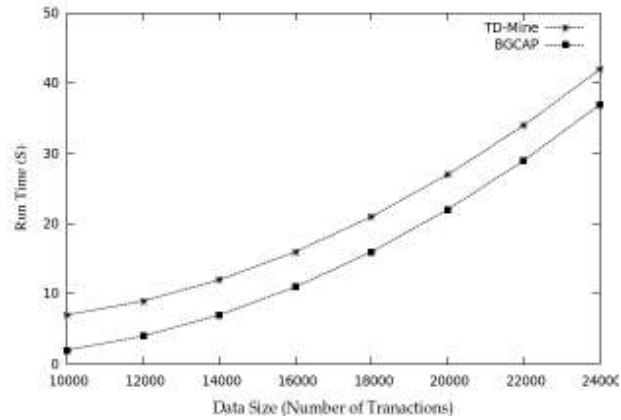

Figure 8: Data Size Vs. Run Time

The minimum support threshold (*minSup*) is another important factor that affects the performance of BGCAP. This is demonstrated in Figure 9 where we show how different support thresholds (for a fixed 2000 transactions) have different execution times. As the data size is fixed, the mining only depends on the *minSup* values. As the *minSup* value increases, the execution time gradually decreases as it would be a smaller number of patterns that are to be mined. In the Figure 9, the graph shows that TD-Mine takes more time to generate patterns compared to BGCAP and they both tend towards the same run time as *minSup* increases. It is observed that the approach adopted by BGCAP reduces the run time by 10-15% than that of TD-Mine.

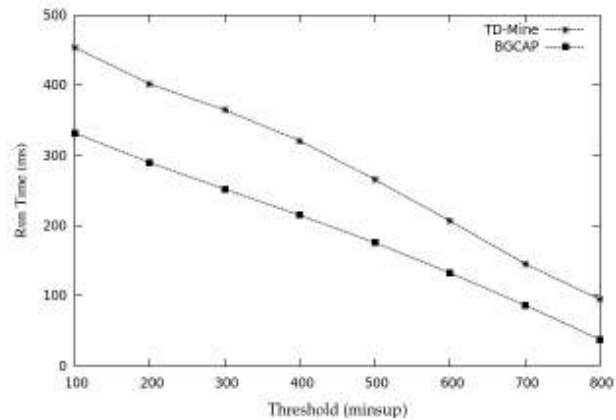

Figure 9: Threshold Vs. Run Time

Next, we mine Web sequential patterns for different thresholds for a fixed data size of 2000 transactions. Figure 10 shows how many patterns can be mined for different support thresholds using BGCAP and TD-Mine. We can see that the number of patterns mined using BGCAP is significantly higher (5-15%) than that of TD-Mine as shown in Figure 10. This is because of the Bidirectional pattern-growth approach adopted by BGCAP that is more scalable and accurate than TD-Mine. However, as the *minSup* increases to the order of 1000 frequent items, the number of patterns generated using both these approaches will tend to be the same as such patterns rarely occur for such large values of threshold.

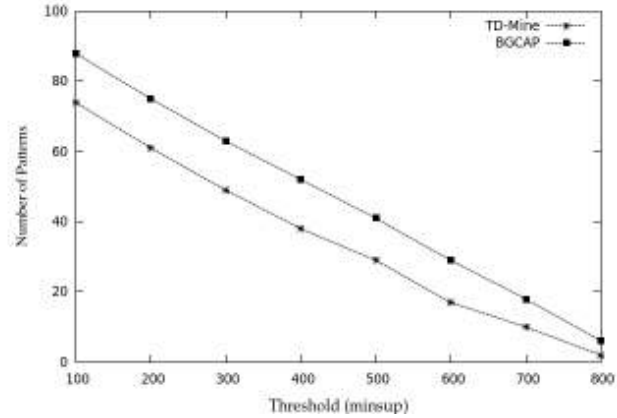

Figure 10: Threshold Vs. Number of Patterns

After generating the Web 2-sequence patterns, BGCAP analyses them using CMA to derive the prefetching rules in terms of periodicity and cyclic behaviour. Consider the same example in Section 5, where we derived the 2-sequence patterns: {<1 3>, <1 6>, <3 6>, <1 7>, <6 7>, <5 6>}. After analysis, we get the following information as shown in Table 7.

Table 7. Example of Prefetching Rules.

| 2-sequence Pattern | Periodicity (s) | Cyclic Behaviour (s) |
|---|---|---|
| <1 3> | 9 | 57 |
| <1 6> | 5 | 93 |
| <3 6> | 7 | 134 |
| <1 7> | 3 | 68 |
| <6 7> | 8 | 74 |
| <5 6> | 4 | 101 |

From the above Table 7, for the 2-sequence pattern <1 3>, the Periodicity is 9 seconds which means Web page (3) will be accessed by a user 9 seconds after Web page (1) has been accessed and so it can be prefetched from the server and stored in the client system before that users actually requests the page (3). As the prefetched Web page has already travelled from the server, the user's perceived delay is insignificant as the page (3) is simply fetched from the client system's memory itself when the user requests the prefetched Web page. The threshold of cyclic behaviour for <1 3> is 57 seconds which means that this behaviour stops after 57 seconds, *i.e.,* the user does not request page (3) after page (1) has been accessed therefore implying that (3) need not be prefetched after the cyclic behaviour has been reached and hence it can be removed from the client's memory. Thus, it reduces network traffic and saves resources by not prefetching Web pages that shall never be requested by the users. Similarly, the prefetching rules can be generated for *n*-sequence Web patterns.

## 7. CONCLUSIONS

The proposed mechanism BGCAP mines Web sequential patterns using UDDAG and analyzes them using CMA to generate Web Prefetching rules. As UDDAG is based on Bidirectional pattern growth, BGCAP performs only (log *n*+1) levels of recursion for mining *n* Web sequential patterns. Prefetching rules generated based on Periodicity and Cyclic Behaviour of 2-sequence Web patterns is very accurate and the said rules hold good only until the threshold of cyclic behaviour has been reached, thus helping to implement a dynamic necessity-based prefetching strategy. Further, our experimental results show that prefetching rules generated using BGCAP is 5-10% faster for different data sizes and 10-15% faster for a fixed data size than TD-Mine. Also, BGCAP generates about 5-15% more prefetching rules than TD-Mine.

Authors

Srikantaiah K C is an Associate Professor in the Department of Computer Science and Engineering at S J B Institute of Technology, Bangalore, India. He obtained his B.E and M.E degrees in Computer Science and Engineering from Bangalore University, Bangalore. He is presently pursuing his Ph.D programme in the area of Web mining in Bangalore University. His reaesrch interest is in the area of Data Mining, Web Mining and Semantic Web.

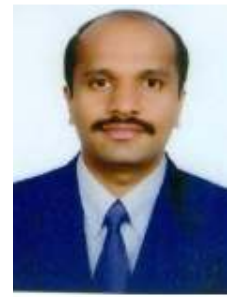

**Krishna Kumar N** received the BE Degree in Information Science and Engineering and MTech. Degree in Computer Engineering from Visvesvaraya Technological University, Belgaum, India, in the years 2010 and 2012, respectively. His research interests include Web Mining, Web Technologies and Computer Networks.

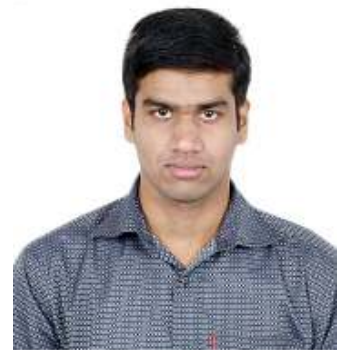

Venugopal K R is currently the Principal, University Visvesvaraya College of Engineering, Bangalore University, Bangalore. He obtained his Bachelor of Engineering from University Visvesvaraya College of Engineering. He received his Masters degree in Computer Science and Automation from Indian Institute of Science Bangalore. He was awarded Ph.D. in Economics from Bangalore University and Ph.D. in Computer Science from Indian Institute of Technology, Madras. He has a distinguished academic career and has degrees in Electronics, Economics, Law, Business Finance, Public Relations, Communications, Industrial Relations, Computer Science and Journalism. He has authored 31 books on Computer Science and Economics, which include Petrodollar and the World Economy, C Aptitude, Mastering C, Microprocessor 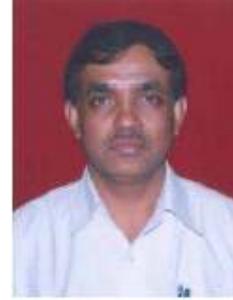 Programming, Mastering C++ and Digital Circuits and Systems etc.. During his three decades of service at UVCE he has over 250 research papers to his credit. His research interests include Computer Networks, Wireless Sensor Networks, Parallel and Distributed Systems, Digital Signal Processing and Data Mining.

L M Patnaik is currently Honorary Professor, Indian Institute of Science, Bangalore. He was a Vice Chancellor, Defense Institute of Advanced Technology, Pune, India. He was a Professor since 1986 with the Department of Computer Science and Automation, Indian Institute of Science, Bangalore. During the past 35 years of his service at the Institute he has over 700 research publications in refereed International Journals and refereed International Conference Proceedings. He is a Fellow of all the four leading Science and Engineering Academies in India; Fellow of the IEEE and the Academy of Science for the Developing World. He has received twenty national and international awards; notable among them is the IEEE Technical Achievement Award for his significant contributions to High Performance Computing and Soft Computing. His areas of research interest have been Parallel and 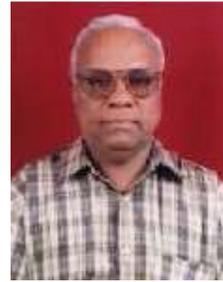 Distributed Computing, Mobile Computing, CAD for VLSI circuits, Soft Computing and Computational Neuroscience.